\documentclass[twocolumn, aps, prl, longbibliography, groupedaddress, superscriptaddress]{revtex4-2}

\usepackage{xcolor}
\usepackage{amsmath}
\usepackage{amssymb}
\usepackage{graphicx}
\usepackage[english]{babel}
\usepackage{textcomp}
\usepackage{braket}
\usepackage{hyperref}
\usepackage{float}
\usepackage{mathtools}
\usepackage{dsfont}
\usepackage{lineno}

\newcommand{\fref}[1]{Fig.~\ref{#1}}

\newcommand{\dgr}{^{\dagger}}
\newcommand{\enum}[1]{\mathit{e}^{#1}}
\newcommand{\mean}[1]{\langle#1\rangle}

\DeclarePairedDelimiter\abs{\lvert}{\rvert}%

\begin{document}
\title{Frustrated magnets without geometrical frustration in bosonic flux ladders}

\author{Luca Barbiero}
\altaffiliation{These authors contributed equally to this work.}
\affiliation{Institute for Condensed Matter Physics and Complex Systems, DISAT, Politecnico di Torino, I-10129 Torino, Italy.}

\author{Josep Cabedo}
\altaffiliation{These authors contributed equally to this work.}
\affiliation{Departament de F\'isica, Universitat Aut\'onoma de Barcelona, E-08193 Bellaterra, Spain}
\email[Electronic address: ]{josep.cabedo@uab.cat}

\author{Maciej Lewenstein}
\affiliation{ICFO - Institut de Ciencies Fotoniques, The Barcelona Institute of Science and Technology,
Av. Carl Friedrich Gauss 3, 08860 Castelldefels (Barcelona), Spain}
\affiliation{ICREA, Pg. Llu\'is Companys 23, 08010 Barcelona, Spain}

\author{Leticia Tarruell}
\affiliation{ICFO - Institut de Ciencies Fotoniques, The Barcelona Institute of Science and Technology,
Av. Carl Friedrich Gauss 3, 08860 Castelldefels (Barcelona), Spain}
\affiliation{ICREA, Pg. Llu\'is Companys 23, 08010 Barcelona, Spain}

\author{Alessio Celi}
\affiliation{Departament de F\'isica, Universitat Aut\'onoma de Barcelona, E-08193 Bellaterra, Spain}

\date{\today}

\begin{abstract}
We propose a scheme to realize a frustrated Bose-Hubbard model with ultracold atoms in an optical lattice that comprises the frustrated spin-1/2 quantum \emph{XX} model. Our approach is based on a square ladder of magnetic flux $\sim\pi$ with one real and one synthetic spin dimension. Although this system does not have geometrical frustration, we show that at low energies it maps into an effective triangular ladder with staggered fluxes for specific values of the synthetic tunneling. We numerically investigate its rich phase diagram and show that it contains bond-ordered-wave and chiral superfluid phases. Our scheme gives access to minimal instances of frustrated magnets \emph{without the need for real geometrical frustration}, in a setup of minimal experimental complexity.
\end{abstract}

\maketitle

\paragraph{Introduction.} The interplay between geometrical frustration and quantum fluctuations leads to exotic states of matter such as resonating valence bond and quantum spin liquid phases \cite{ANDERSON_1973_RVB, balents-2010-spin-liquids, Mila-Book-2011}. The simplest models encompassing the richness of frustrated quantum magnets are antiferromagnetic Heisenberg Hamiltonians on triangular lattices \cite{Diep-Book-frustrated-spins}, which include deconfined quantum critical points \cite{senthil-science-2004-deconfined, Sandvik_prl_2007_deconfined, Jiang_prb_2019}, anyonic liquids  \cite{Rahmani-prl-2014}, and where spontaneous dimerization and chiral order appear \cite{Majumdar-1969,Haldane-prb-1982,Okamoto-pla-1992, White-prb-1996,Nersesyan-prl-1998}. In this Letter, we focus on a minimal instance of the frustrated antiferromagnetic Heisenberg model, the spin-1/2 quantum \emph{XX} model on a triangular two-leg ladder. We show that its rich phase diagram can be effectively accessed with ultracold bosons in flux ladders of square lattice geometry, a setup of minimal experimental complexity that is routinely realized 
with real-space \cite{Atala_2014, Tai_2017} and synthetic dimension approaches 
\cite{Celi-2014, Mancini-15, Stuhl-15, Livi-16, Kolkowitz-17, Han_prl_2019_synth_tube, Li-2022, Anderson-2020, Zhou_science_2023_synth_ladder_Hall_resp}.

While the investigation of spin-$1/2$ Heisenberg triangular ladder systems in solid state materials is a very active field of research \cite{Drechsler-prl-2007,Graaf-prb-2002,Wolter-prb-2012,Grafe-sr-2017,Orlova-prl-2017,Ueda-prb-2020, Grams-2022}, the broad tunability of ultracold atoms offers an attractive alternative to investigate magnetic frustration in a pristine setting and gives access to new observables. On the one hand, Fermi gases in triangular optical lattices or optical tweezer arrays provide ideal implementations of the celebrated $J_1$-$J_2$ antiferromagnetic Heisenberg Hamiltonian \cite{Tarruell-nature-2012, Yang-2021, Yan-2022}, although achieving experimental temperatures below the superexchange energy scale remains a formidable challenge \cite{Mazurenko_Nature_2017_hubbard_magnetic_order, Mongkolkiattichai_2022}. On the other hand, strongly interacting bosonic systems subjected to artificial magnetic fluxes also display frustrated magnetic phases, but at larger (and accessible) energy scales set by the tunneling \cite{Eckardt-2010-epl, Celi-2016}. However, despite tremendous progress in the realization of artificial gauge fields in such systems using real-space lattices \cite{Struck-science-2011, Struck-2013, Aidelsburger-2011, Aidelsburger-2013, Miyake-2013, Atala_2014, Tai_2017}, the combination of large magnetic fluxes and strong interactions leads to detrimental heating processes that hinder the investigation of quantum magnetism \cite{weitenberg_natphys_2021_floquet}.

An alternative approach, more resilient to heating, is to employ semisynthetic flux ladders with one fictitious dimension constituted by internal spin states coupled via two-photon Raman transitions \cite{Boada-12, Celi-2014}, a system that has been successfully employed to experimentally investigate few-leg square ladder systems in both noninteracting \cite{Mancini-15, Stuhl-15, Livi-16, Kolkowitz-17} and strongly interacting regimes  \cite{Zhou_science_2023_synth_ladder_Hall_resp}, but is not straightforward to generalize to triangular geometries without introducing additional heating mechanisms \cite{Anisimovas-pra-2016, Suszalski-pra-2016}. In this Letter, we propose a scheme to realize a frustrated quantum spin model---inbuilt in a semisynthetic flux ladder---\emph{without the need for explicit geometrical frustration}. Building on our previous work \cite{Cabedo_epjd_2020}, we exploit an analytical map between the square flux ladder at low energies and a triangular system with staggered magnetic fluxes. Related schemes have been very recently proposed in fermionic synthetic ladders \cite{Mamaev-2021, Mamaev_2022}, although no effective frustration was investigated in this setting. Here, we identify parameter regimes where the effective geometric frustration plays an important role at accessible temperatures, numerically show that the system displays the same ground state phases as the frustrated spin-1/2 quantum \emph{XX} model \cite{Sato-mplb-2011}, and identify suitable observables to reveal them in current experiments with ultracold atoms.

\begin{figure}[t!]
\centering
\includegraphics[width=0.99\linewidth]{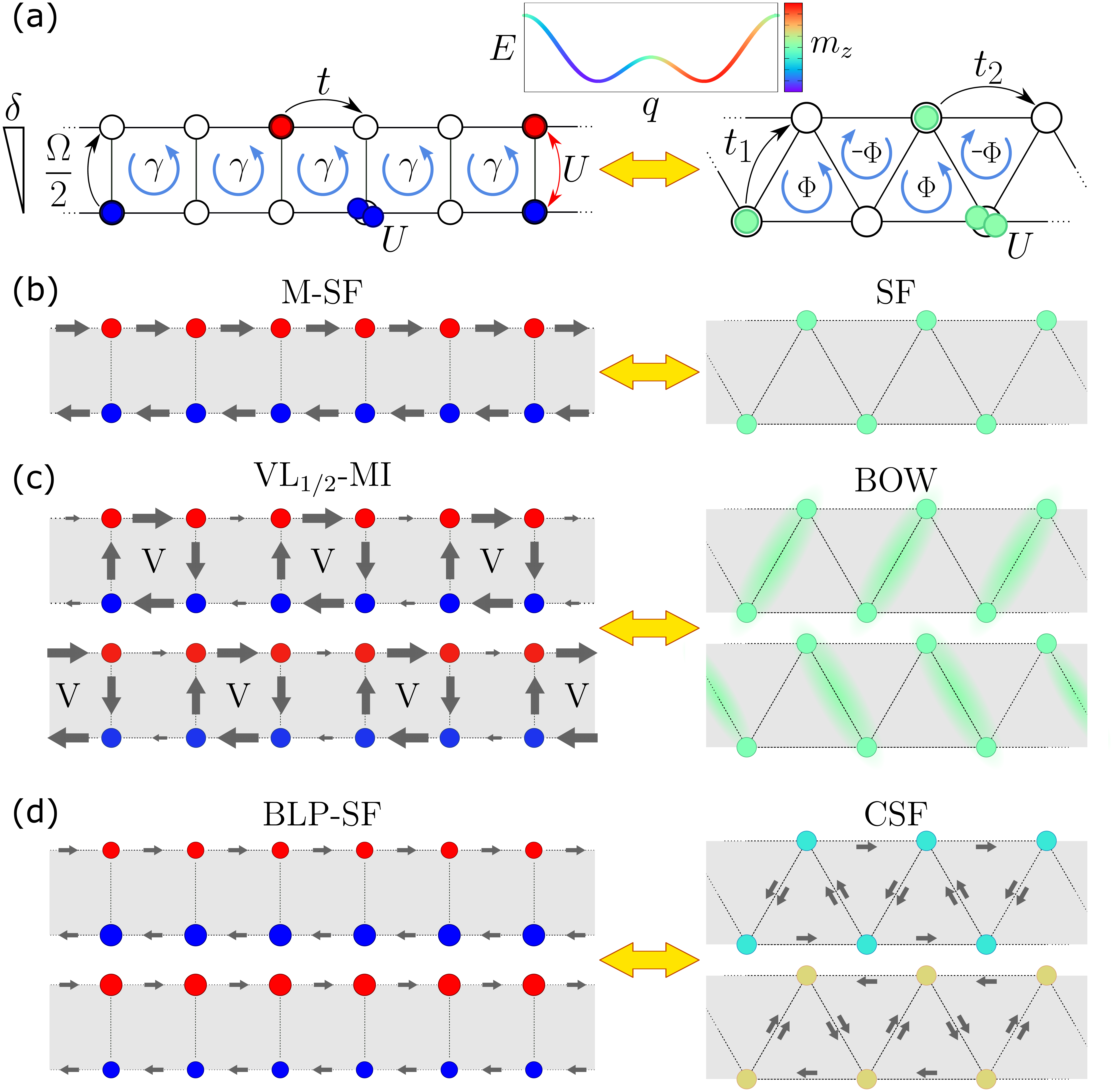}
\caption{Mapping semisynthetic square ladders into frustrated triangular ladders. (a) Left: Original square flux-ladder model \eqref{eq_square_ham}, with intra- and inter-leg tunnelings $t$ and $\Omega/2$, inter-leg offset $\delta$, on- and off-site interactions $U$, and magnetic flux $\gamma$; Right: Truncated lower-band Hamiltonian \eqref{eq_triang_Hamiltonian}, describing an effective triangular ladder with complex tunnelings $t_1$ and $t_2$, on-site interactions $U$, and staggered flux $\Phi$. Inset: Single-particle dispersion relation of the lower energy band of the square flux ladder for strong $\Omega$ and $\gamma \sim \pi$. (b)--(d) Current (arrow) and density (circle) patterns of the flux ladder phases at average density $\rho = 1/4$ (left), and corresponding phases of the effective triangular model at half filling (right). (b) Meissner superfluid (M-SF) -- superfluid (SF) of the triangular model. (c) Vortex lattice insulator (VL$_{1/2}$-MI) -- bond-ordered wave (BOW). (d) Biased-ladder superfluid phase (BLP) -- chiral superfluid (CSF). Circles and arrows in the square ladder sketches are scaled according to their numerical values at $ \Omega =10 t$ and $\delta=0$, with $\gamma$ and $U$ adjusted so that $t_2/\vert t_1 \vert = 0.2$ in (b), $t_2/\vert t_1 \vert = 0.5$ in (c) and $t_2/\vert t_1 \vert = 1.0$ in (d), and $U= 10 \vert t_1 \vert$. The BLP is signaled by the difference of circle size between both legs. Color scale: Spin composition $m_z$ of the ground states in the semisynthetic ladder.}
\label{Fig_triang_map}
\end{figure}

\paragraph{Model.}
We consider the two-leg semisynthetic bosonic flux ladder \cite{Celi-2014} in the translation-invariant gauge-transformed frame \cite{Supplemental_Mat_triang_ladder}, see Fig. \ref{Fig_triang_map}(a), 
\begin{align}\label{eq_square_ham}
H_{\square}=&\sum_{j, \sigma}\big(-t e^{-\imath\gamma\sigma}a_{j+1,\sigma}^\dagger +\frac{\Omega}{4} a_{j,-\sigma}^\dagger + \frac{\delta}2\sigma a_{j,\sigma}^\dagger \big)a_{j,\sigma}  \! + \! \text{H.c.}
\nonumber\\
+&\sum_{j, \sigma}\left( \frac{U_{\sigma,\sigma}}{2}n_{j,\sigma}(n_{j,\sigma}\!-1)+\frac{U_{\sigma,-\sigma}}{2}n_{j,\sigma}n_{j,-\sigma} \right).
\end{align}
Here, $a_{i\sigma}^\dagger (a_{i\sigma})$ creates (annihilates) a boson in site $i$ of a one-dimensional (1D) optical lattice of length $L$---we restrict ourselves to 1D systems throughout this Letter---and two different internal atomic states $\sigma=\pm1/2$ coupled via Raman transitions realize the two legs of the ladder. The intra-leg tunneling $t$ is the conventional tunneling rate along the lattice, while the inter-leg tunneling amplitude is proportional to the Raman Rabi frequency $\Omega$. The two-photon Raman detuning yields a potential interleg offset $\delta$, and a classical magnetic flux $\gamma$ results from the momentum transfer of the Raman beams.  Atoms experience intra-leg on-site and inter-leg nearest-neighbor interactions, which we choose to be identical $U_{\sigma,\sigma}=U_{\sigma,-\sigma}=U$. The total atom number is $N=\sum_{i=1}^{L}\sum_\sigma n_{i,\sigma}$, with $n_{i,\sigma}=a_{i,\sigma}^\dagger a_{i,\sigma}$ the density in site $i$. 

In the strong Raman coupling limit $\Omega \gg t$, the two single-particle dispersion bands of \eqref{eq_square_ham} are separated by an energy gap $\sim \Omega$ \cite{Supplemental_Mat_triang_ladder}; we denote them as lower- and higher-band dressed states. When the condition $\Omega\gg U$ is also fulfilled, the low-energy properties of the system are captured by a Hamiltonian that includes only lower-band modes, \begin{equation}
H_{\bigtriangleup}=\sum_{l=1,2}t_l\sum_{i}(b_i^\dagger b_{i+l}+\text{H.c.})+\frac{U}{2}\sum_{i}\tilde{n}_i(\tilde{n}_i-1),
\label{eq_triang_Hamiltonian}
\end{equation}
where $b_j^\dagger (b_j)$ are the bosonic creation (annihilation) operators for the inverse-Fourier-transformed lower-band dressed states and $\tilde{n}_j= b_j^\dagger b_j$ [see Supplemental Material \cite{Supplemental_Mat_triang_ladder} for the derivation]. In this regime, the effective Hamiltonian \eqref{eq_triang_Hamiltonian} describes a system of $N$ bosons in a lattice of length $L$ with the same on-site interaction $U$ but with effective nearest neighbor (NN) $t_1$ and next-nearest neighbor (NNN) $t_2$ complex tunnelings. It is thus equivalent to a triangular ladder with 
staggered magnetic flux $\Phi = \pi -2\delta\tan(\gamma/2)/\Omega +\mathcal{O}\left[(\delta/\Omega)^2\right]$ [see Fig. \ref{Fig_triang_map}(a)]. At $\delta=0$, the staggered flux ladder is fully frustrated with $\Phi = \pi$. To order $\mathcal{O}\left[(t/\Omega)^2\right]$, the effective tunneling amplitudes in $H_{\bigtriangleup}$ relate to the parameters of $H_{\square}$ by \cite{Supplemental_Mat_triang_ladder}
\begin{equation}
t_1 \simeq -t\cos(\gamma/2) \quad\text{and}\quad t_2 \simeq t^2\sin^2(\gamma/2)/\Omega.  \label{eq_t1t2}
\end{equation}
In the ultracold atom context, the triangular Bose-Hubbard ladder Hamiltonian \eqref{eq_triang_Hamiltonian} has been mainly studied at unity filling \cite{Greschner-pra-2013,Zaletel-prb-2014, Romen_PhysRevB_2018_CMI} and low densities \cite{Greschner-prb-2019} (for nonstaggered fluxes, see Ref. \cite{Halati_arXiv_2022}). Moreover, detailed studies of the hardcore boson (HCB) limit ($U\rightarrow\infty$) at half filling $N/L=1/2$ and flux $\phi = \pi$, where the system is further mapped to a frustrated spin-$1/2$ quantum \emph{XX} model ($b_{i}\dgr, b_j \rightarrow S_{i}^+, S_{j}^-$, see SM \cite{Supplemental_Mat_triang_ladder}) have been performed \cite{Sato-mplb-2011, Mishra-2013, Mishra-2014}. In this regime, a gapless superfluid (SF) is found at low values of $\abs{t_2/t_1}$, signaled by the  power-law decay of the one-body correlator $g^1(|i-j|)=\langle b^\dagger_ib_j\rangle$. At intermediate values of $\abs{t_2/t_1}$, a gapped translation-breaking bond-ordered-wave (BOW) phase is stabilized [see Fig. \ref{Fig_triang_map}(c)]. It is signaled by nonzero values of the two-point operator $O_{\text{BO}}=\sum_i [(-1)^i/L]( b^\dagger_ib_{i+1}+b_ib^\dagger_{i+1})$. Its insulating nature is signaled by the exponential decay of $g^1(|i-j|)$ and by a finite charge gap  $\Delta_c = E_{L,N+1}+E_{L,N-1} - 2 E_{L,N}$, computed from the extrapolated groundstate energies at $N=L/2$. Finally, for larger $\abs{t_2/t_1}$, a gapless chiral superfluid (CSF) phase emerges. There, the system presents two nonequivalent minima in the dispersion relation, and interatomic interactions favor the occupation of either of the two minima. This degeneracy yields two solutions that spontaneously break a $Z_2$ parity symmetry [see Fig. \ref{Fig_triang_map}(d)], and exhibit a finite chirality $k_i=2\imath(b_ib_{i+1}^\dagger-b_i^\dagger b_{i+1})$. Thus, the chiral correlation function $k^2(|i-j|)=\langle k_i k_j\rangle$ identifies the CSF.

In our realization of the triangular model \eqref{eq_triang_Hamiltonian}, we can widely adjust the ratio $\abs{t_2/t_1}$ while arbitrarily approaching the HCB limit within the effective model by setting $\abs{t_{1,2}} \ll U \ll \Omega$. Therefore, we expect that at filling $\rho=1/4$ the flux ladder Hamiltonian \eqref{eq_square_ham} reproduces the whole phase diagram of the quantum \emph{XX} model with the order parameters written in terms of the currents and densities of the undressed bosons, and that the phases are experimentally accessible \cite{Supplemental_Mat_triang_ladder}.

The SF phase of the triangular model translates into the Meissner superfluid phase (M-SF) of the flux ladder \cite{Orignac-prb-2001} [see \fref{Fig_triang_map}(b)], characterized by vanishing rung currents and off-diagonal quasi-long range order. The BOW phase corresponds to a vortex lattice insulating phase (VL$_{1/2}$-MI) of maximal vortex filling $\rho_{v} = 1/2$, where the effective dimers correspond to the vortex plaquettes [see \fref{Fig_triang_map}(c)] The nature of the BOW phase in the bare basis is easily understood from the susceptibility of the energy against the explicit dimerization of the leg tunnelings \cite{Supplemental_Mat_triang_ladder}, through which one can identify 
\begin{equation}\label{eq_OBO_bare}
O_\mathrm{BO} \simeq   \sum_{j,\sigma} \frac{2(-1)^j \operatorname{Re}(te^{-\imath\gamma\sigma} a_{j+1,\sigma}^\dagger a_{j,\sigma})}{L t \cos(\gamma/2)}.
\end{equation}
The BOW phase is characterized by the staggered current patterns of the vortices, and is signaled by the staggered leg current
\begin{equation}\label{eq_jsl}
j_{sl} = \frac{1}{L} \sum_{j,\sigma}  4 \sigma (-1)^j  \operatorname{Im}\left( te^{-\imath\gamma\sigma} a_{j+1,\sigma}^\dagger a_{j,\sigma}\right),
\end{equation}
together with the exponential decay of the one-body correlator. Finally, the CSF corresponds to a biased-ladder superfluid phase (BLP-SF), characterized by a spontaneous density imbalance between both legs of the ladder \cite{Wei_pra_BLP_theory, Tokuno-njp-2014} [see \fref{Fig_triang_map}(c)]. Around  $\delta=0$,
\begin{equation}\label{eq_kj_bare}
k_j \simeq -\frac{2\Omega}{t\sin(\gamma/2)}m_z^{(j)},
\end{equation}
where $m_z^{(j)}=\sum_{\sigma}\sigma a_{j,\sigma}^\dagger a_{j,\sigma}$ is the magnetization or the inter-leg population imbalance at site $j$ \cite{Supplemental_Mat_triang_ladder}.    

Based on these correspondences, we propose to detect the different phases by measuring the response of the system to the explicit breaking of the $Z_2$ parity and translation symmetries of the Hamiltonian. The former is easily  achieved by setting the Raman detuning $\delta$ to nonzero values. The latter can be implemented by a spatial modulation of the optical lattice using superlattice potentials to dimerize the lattice structure, parameterized by including in \eqref{eq_square_ham} a position dependent tunneling strength $t_j = t(1+\Delta(-1)^j)$ between sites $j$ and $j+1$. In this way, the spontaneous breaking of the $Z_2$ symmetry of the ground state in the CSF (or BLP-SF) phase is signaled by the discontinuity in the mean magnetization $\mean{m_z} = \frac{1}{N}\sum_{j} \mean{m_z^{(j)}}$ around $\delta = 0$. Similarly, the spontaneous dimerization of the BOW (or V$_{1/2}$-MI) phase is signaled by the jump of the staggered currents \eqref{eq_jsl} around $\Delta = 0$.

\begin{figure}[t!]
\centering
\includegraphics[width=0.95\linewidth]{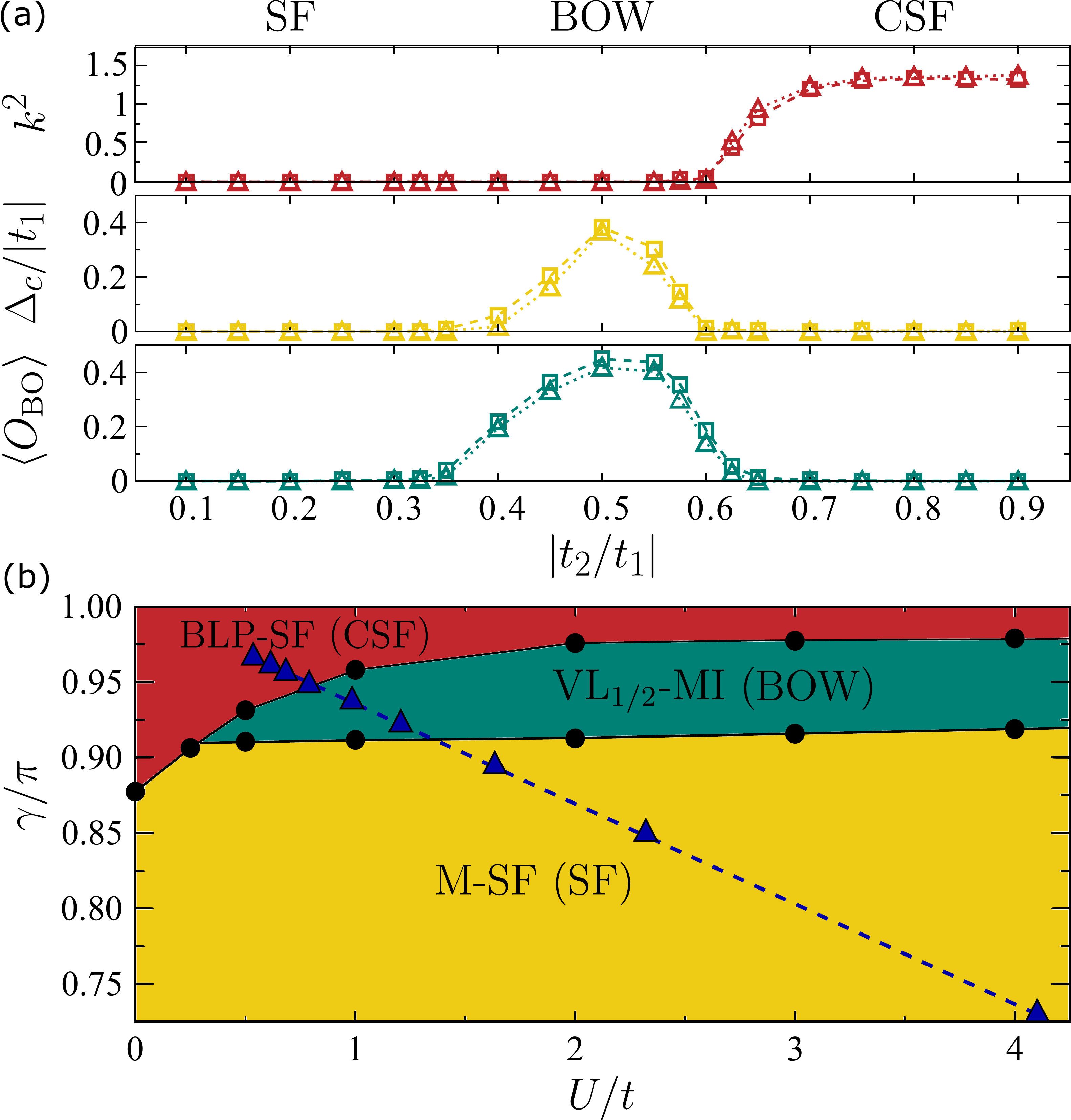}
\caption{Phase diagram. (a) DMRG-computed values of chiral correlation function $k^2(L/2)$ (top panel), charge gap $\Delta_c$ (central panel), and two-point bond-order operator $O_{\text{BO}}$ (bottom panel) as a function of $|t_2/t_1|$ extrapolated to $L \rightarrow \infty$. Triangles: Effective triangular staggered-flux ladder \eqref{eq_triang_Hamiltonian} at half filling, with $U  = 10 \vert t_1 \vert$ and $\Phi = \pi$. Squares: Square flux ladder \eqref{eq_square_ham} at $\rho = 1/4$ and $\delta = 0$, where we have fixed $\Omega = 20 t$ and adjusted $\gamma$ and $U$ to match the corresponding values of $t_2/t_1$ and $U/\lvert t_1\rvert$. All quantities are extracted via finite size extrapolation using system lengths up to $L = 120$ ($L = 80$) for the triangular (square) ladder model. Only the central half of the sites is used to compute the expected values. (b) Phase diagram of the semisynthetic square flux ladder \eqref{eq_square_ham} at $\rho = 1/4$ for $\Omega =20 t$ and $\delta = 0$. Black circles: Phase boundary from DMRG simulations. Blue triangles:  Set of points where $U= 10 t \cos(\gamma/2)=10\lvert t_1\rvert$ and the effective triangular ladder \eqref{eq_triang_Hamiltonian} tunneling ratio $\vert t_2/t_1\vert$ takes values from 0.9 to 0.1, corresponding to the curves of (a).}
\label{Fig_phase_diag}
\end{figure}

\paragraph{Numerical results.}

To assess the predictions of the effective model \eqref{eq_triang_Hamiltonian}, we run density-matrix-renormalization-group (DMRG) \cite{White-prl-1992, Itensor} simulations of Hamiltonian \eqref{eq_square_ham} in the regimes discussed. In \fref{Fig_phase_diag}(a), we show the values of $\Delta_c$, $O_{\text{BO}}$, and $k^2$ as a function of $\vert t_1/t_2 \vert$, for the ground states of both the effective triangular Hamiltonian \eqref{eq_triang_Hamiltonian} at half filling (triangles) and the original square ladder Hamiltonian \eqref{eq_square_ham} at filling $\rho = 1/4$ (squares). For the latter, we set $\Omega = 20 t$, use \eqref{eq_t1t2} to retrieve the corresponding values of $t_2$ and $t_1$, and use expressions \eqref{eq_OBO_bare} and \eqref{eq_kj_bare} to compute $O_{\text{BO}}$ and $k^2$. In both cases, the interaction strength is set to $U = 10 t \cos(\gamma/2)=10 \vert t_1 \vert$, which realizes the strongly interacting regime of the effective triangular ladder. We observe very good agreement between both models in this regime of parameters. Moreover, the phase diagram predicted in the HCB limit of the effective model \cite{Sato-mplb-2011} is preserved for large but finite values of $U/\vert t_1 \vert$ and $\Omega$. The gapless SF (M-SF) phase appears for sufficiently low $\vert t_2/t_1 \vert$. For higher values of  $\vert t_2/t_1\vert$ the quasi-long range order is lost and a phase with gap $\Delta_c \neq 0$ occurs instead. It is captured by finite values of $O_{\text{BO}}$, signaling the dimerization that characterizes the BOW phase of the triangular model, or, equivalently, the vortex structures of the V$_{1/2}$-MI phase in the square ladder (see \eqref{eq_OBO_bare}). Finally, a gapless phase analogous to the CSF of the spin chain appears when $\abs{t_2/t_1}$ is further increased. It is characterized by the long-range order of $k^2$, signaling the inter-leg population imbalance of the BLP-SF phase (see \eqref{eq_kj_bare}).

\fref{Fig_phase_diag}(b) shows the phase diagram of the square flux ladder in the $U$-$\gamma$ plane for $\Omega = 20t$. The blue triangles fix the condition $U=10 t\cos{\gamma/2}=10\vert t_1 \vert$, and correspond to the curves displayed in \fref{Fig_phase_diag}(a). Remarkably, the three phases predicted by the effective frustrated \emph{XX} model persist in wide regions of parameter space. For stronger interactions---where the truncation to the lower band is no longer accurate---a charge density wave phase (not displayed here) has been shown to appear \cite{Piraud-2015, Greschner-2016}, while the M-SF survives to the HCB limit of the flux ladder at quarter filling \cite{Petrescu-13}.

\begin{figure}[b!]
\centering
\includegraphics[width=0.95\linewidth]{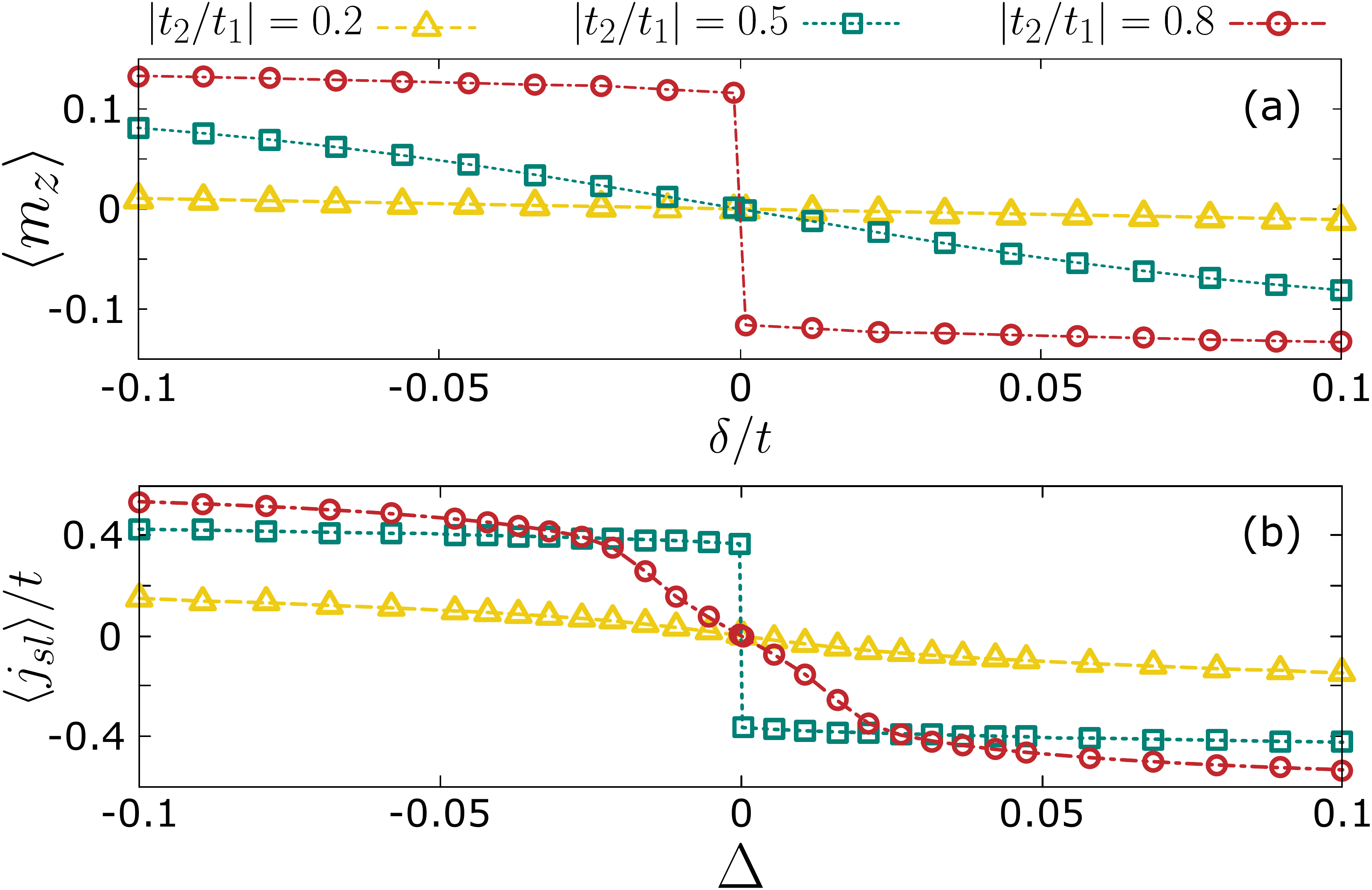}
\caption{(a) Magnetization $m_z$ as a function of the Raman detuning $\delta$, for the ground state of Hamiltonian \eqref{eq_square_ham} at $\rho = 1/4$, with $\Omega = 10 t$ and adjusting $\gamma$ and $U$ so that $U = 10 \vert t_1 \vert$ and $\vert t_2/ t_1\vert = 0.2$ (yellow triangles), $0.5$ (teal squares),  $0.8$ (dark red circles), with $t_1 < 0$. (b) Staggered leg current $j_{sl}$ defined in \eqref{eq_jsl} as a function of the lattice dimerization $\Delta$. All quantities are extrapolated to the thermodynamic limit by considering system sizes up to $L= 80$.}
\label{Fig_mz_jsl}
\end{figure}

Finally, we numerically assess the protocol described above to probe the three phases in the square flux ladder. Figure \ref{Fig_mz_jsl}(a) shows the mean magnetization $m_z$ of the ground state as a function of $\delta$ for $\Omega=10t$. We adjust $\gamma$ to different values of $\vert t_2/t_1 \vert$ and keep $U=10 \vert t_1 \vert$. For $\vert t_2/t_1 \vert = 0.8$, the discontinuity in $m_z$ around $\delta = 0$ signals the spontaneous breaking of the $Z_2$ parity symmetry that characterizes the BLP-SF (CSF) phase. Similarly, \fref{Fig_mz_jsl}(b) shows the expected value of the staggered leg current $j_{sl}$ [see Eq.\eqref{eq_jsl}] as a function of $\Delta$. As expected, for $\vert t_2/t_1 \vert = 0.5$, the spontaneous dimerization that characterizes the VL$_{1/2}$-MI (BOW) is signaled by the discontinuity of $j_{sl}$ around $\Delta = 0$.

\paragraph{Experimental implementation.} The flux-ladder model discussed above can be implemented using a Raman-coupled Bose gas in a 1D optical lattice. This approach allows realizing large rung couplings along the synthetic direction with low Raman intensities that ensure small heating rates \cite{Stuhl-15}. For concreteness, we focus on $^{41}$K atoms at large magnetic fields, where the low sensitivity of $\delta$ to magnetic field fluctuations \cite{Buser_2020} provides a fine control of the effective staggered flux $\Phi$. However, our scheme can be implemented equivalently with $^{87}$Rb atoms using dynamical decoupling schemes \cite{Trypogeorgos_2018, Anderson_2018}. We consider a blue-detuned retro-reflected 1D lattice of wavelength $\lambda_{L,s}=532$ nm of depth $5E_{L,s}$ for the physics, and an additional retro-reflected lattice of wavelength $\lambda_{L,l}=1064$ nm overlapped with it for preparing the required $\rho=1/4$ filling and for dimerizing the system in the symmetry breaking measurements of \fref{Fig_mz_jsl} \cite{Supplemental_Mat_triang_ladder}. Using two Raman beams at the tune-out wavelength $769$ nm and forming a $\sim45^{\circ}$ angle with the lattice beams yields the required flux $\gamma\sim\pi$ \cite{Celi-2014}. In such conditions, the BOW and CSF regimes of the effective triangular model can be realized with tunneling rates $\vert t_{1,2} \vert /h$ above $100$ Hz, while reaching the strongly interacting regime $U=10|t_1|$ by confining the gas along the transverse directions with two $1064$-nm perpendicular optical lattices of depth $\sim45E_{L,l}$. Moreover, the Raman coupling $\Omega \sim 10t$ remains sufficiently large to satisfy the lower-band approximation, while enabling gas lifetimes much larger than the tunneling timescales of the effective model \cite{Supplemental_Mat_triang_ladder}. Finally, the magnetization and staggered leg currents of \fref{Fig_mz_jsl} can be simply determined by measuring the atomic spin populations and leg currents using a combination of Stern-Gerlach, superlattice, and time-of-flight techniques \cite{Stuhl-15, Atala_2014}. Thus, our proposal is immediately accessible in current experiments.

\paragraph{Conclusion.} We have shown that semisynthetic flux ladders enable the experimental realization of a frustrated \emph{XX} Heisenberg model with bosons in lattices without real geometric frustration. Our scheme is robust and does not suffer from the heating limitations of Floquet-engineering or higher-band schemes \cite{Struck-science-2011, Struck-2013, Wang_arXiv_2022_porbital_triang}. Its tunability makes it ideal for detecting multipartite entanglement \cite{Singha_Roy_prb_2022_frutrated_BHM}, and for studying transport, out-of-equilibrium and thermal \cite{Buser_PRA_2019_temperature} properties, which are challenging for classical numerical simulations but start to be accessible experimentally \cite{Zhou_science_2023_synth_ladder_Hall_resp}. Whether such effective low-energy dressed-state descriptions can be successfully applied to spin models implemented in other experimental platforms \cite{Leseleuc_Science_2019_rydbergs, Lienhard_prx_2020_density_gauge, Grass_pra_2015_synth_flux_ions, Shapira_arXiv_2022_flux_ion_chain_exp, Gorshkov_prl_2011_mol, Bohn_science_2017_mol,  Douglas_Nature_2015_phot_crys,Chang_revmodphys_2018_photon_lat, Colella_2022}, or extended to engineer 2D frustrated models with \cite{Celi-2016} or without \cite{Diliberto-2023-prr} geometric frustration is an open question worth investigating.

\begin{acknowledgements}
\paragraph{Acknowledgements.} We are grateful to C. S. Chisholm and A. Rubio-Abadal for a critical reading of the manuscript.  L. B. acknowledges M. Aidelsburger, N. Baldelli and C. R. Cabrera for discussions on related topics. Research at ICFO was supported by EU Quantum Flagship (PASQuanS2.1, Grant No. 101113690), Spanish MCIN/AEI/10.13039/501100011033 Severo Ochoa program for Centres of Excellence in R\&D (CEX2019-000910-S/10.13039/501100011033) and MCIN Recovery, Transformation and Resilience Plan with funding from European Union NextGenerationEU (PRTR C17.I1), Fundació Privada Cellex, Fundació Mir-Puig, and Generalitat de Catalunya (CERCA program). L. B. acknowledges funding from Politecnico di Torino, starting package Grant No. 54 RSG21BL01 and from the Italian MUR (PRIN “DiQut” grant No.  2022523NA7). L. B. and M. L. acknowledge funding from European Union (ERC AdG-833801 NOQIA, EU Horizon 2020 FET-OPEN OPTO-Logic (Grant No. 899794), and EU Horizon Europe NeQST (Grant No. 101080086)), Spanish AEI/MCIN/10.13039/501100011033 (FIDEUA PID2019-106901GB-I00, PGC2018-097027-B-I00, and QUSPIN RTC2019-007196-7), EU QUANTERA MAQS (funded by State Research Agency MICN/AEI/10.13039/501100011033, PCI2019-111828-2), Generalitat de Catalunya (AGAUR 2021 SGR 01452), Barcelona Supercomputing Center MareNostrum (FI-2022-1-0042 and FI-2023-1-0013), and the Polish National Science Centre (Symfonia Grant No. 2016/20/W/ST4/00314). L. T. acknowledges funding from European Union (ERC CoG-101003295 SuperComp), Spanish MCIN/AEI/10.13039/501100011033 (LIGAS PID2020-112687GB-C21), Deutsche Forschungsgemeinschaft (Research Unit FOR2414, Project No. 277974659), and Generalitat de Catalunya (AGAUR 2021 SGR 01448). Research at UAB was supported by MCIN/AEI/10.13039/501100011033 (LIGAS PID2020-112687GB-C22) and Generalitat de Catalunya (AGAUR 2021 SGR 00138). A. C. acknowledges support from the UAB Talent Research program. All authors acknowledge funding from EU QUANTERA DYNAMITE (funded by MICN/AEI/10.13039/501100011033 and by the European Union NextGenerationEU/PRTR PCI2022-132919 (Grant No. 101017733)), from Generalitat de Catalunya (QuantumCAT U16-011424, co-funded by ERDF Operational Program of Catalonia 2014-2020), from the Ministry of Economic Affairs and Digital Transformation of the Spanish Government through the QUANTUM ENIA project call – Quantum Spain project, and by the European Union through the Recovery, Transformation and Resilience Plan – NextGenerationEU within the framework of the Digital Spain 2026 Agenda. 
\end{acknowledgements}


%

\clearpage
\widetext
\begin{center}
\textbf{\large Supplemental Material}
\end{center}
\setcounter{equation}{0}
\setcounter{figure}{0}
\setcounter{table}{0}
\makeatletter
\renewcommand{\theequation}{S\arabic{equation}}
\renewcommand{\thefigure}{S\arabic{figure}}
\renewcommand{\bibnumfmt}[1]{[S#1]}
\renewcommand{\citenumfont}[1]{S#1}

In this Supplemental Material we include the detailed derivation of the lower-band Hamiltonian (2) presented in the main text. Furthermore, we discuss the nature of the symmetry-broken phases of the effective model back in the original undressed basis, and discuss the meaningful observables of the effective frustrated model in terms of the original square flux ladder. Finally, we briefly assess the experimental viability of our proposal.

\section{System}
We consider a 1D ultracold Bose gas which is subject to Raman dressing, and consider that only two internal states (effective spin DOF) are nearly-resonantly coupled by the Raman transitions. The Raman beams of wavelength $\lambda_R$ intersect with an opening angle $\theta_R$. We arbitrarily set the two-photon recoil momentum along $\hat{x}$, and label the effective single-photon recoil momentum by $k_R = 2\pi \cos\theta_R /\lambda_R$. Furthermore, we consider the dressed gas to be loaded in a 1D optical lattice set along $\hat{x}$, with the potential being deep enough to consider the tight-binding approximation. The lattice is characterized by a laser wavelength $\lambda_L$ and an opening angle $\theta_L$, which define the intersite spacing $d = \pi/k_L$, with $k_L = 2\pi \cos\theta_L/\lambda_L$. Such a configuration can be interpreted as a two-leg semisynthetic bosonic flux ladder \cite{Celi-2014_sup}, described by
\begin{align}\label{Seq_square_ham_nonrotated}
H_{\square}= &\sum_{j}\sum_{\sigma=\pm 1/2}\left(-t \bar a_{j+1,\sigma}^\dagger  +\frac{\Omega}{2}e^{-\imath\gamma j} \bar a_{j,\sigma-1}^\dagger + \frac{\delta}2 \sigma \bar{a}_{j,\sigma}^\dagger \right) \bar a_{j,\sigma} + \text{H.c.} \cr 
+ &\sum_{j}\sum_{\sigma=\pm 1/2}\left( \frac{U_{\sigma,\sigma}}{2}n_{j,\sigma}(n_{j,\sigma}-1)+\frac{U_{\sigma,-\sigma}}{2}n_{j,\sigma}n_{j,-\sigma} \right).
\end{align}
Here $\bar a_{j\sigma}^\dagger ( \bar a_{j\sigma})$ creates (annihilates) a boson in $j_{\mathrm{th}}$ site, with internal atomic state $\sigma$. The two internal states $\sigma=\pm 1/2$ realize the two legs of the ladder. The intra-leg hopping $t$ is the conventional nearest-neighbor (NN) tunneling rate along the lattice. Finally, the inter-leg hopping is provided by the Raman coupling, and it is proportional to the Raman Rabi frequency $\Omega$, and the two-photon Raman detuning yields a potential offset $\delta$ between the two legs. Due to the nature of Raman transitions, atoms tunneling between the legs also pick-up a position-dependent phase, adding up to a nonzero flux $\gamma = 2 k_R d = 2\pi \frac{k_R}{k_L}$ following a close loop around each plaquette. Finally, the atoms experience density-density intra-leg on-site and inter-leg nearest-neighbor interactions of strength $U_{\sigma,\sigma^{\prime}}$, which may depend on the different spin-spin scattering lengths. Here, $n_{j,\sigma}= \bar a_{j,\sigma}^\dagger \bar a_{j,\sigma}$ is the density in site $j,\sigma$.

The explicit dependence of the phase terms on the lattice site can be eliminated by applying the gauge transformation $a_{j,\sigma} = \enum{-i\sigma j \gamma} \bar a_{j,\sigma}$, yielding
\begin{align}\label{Seq_square_ham}
H_{\square}= &\sum_{j}\sum_{\sigma=\pm 1/2}\left(-t e^{-\imath\gamma \sigma} a_{j+1,\sigma}^\dagger +\frac{\Omega}{4} a_{j,-\sigma}^\dagger + \frac{\delta}2 \sigma  a_{j,\sigma}^\dagger \right) a_{j,\sigma} + \text{H.c.} \cr 
+ &\sum_{j}\sum_{\sigma=\pm 1/2}\left( \frac{U_{\sigma,\sigma}}{2}n_{j,\sigma}(n_{j,\sigma}-1)+\frac{U_{\sigma,-\sigma}}{2}n_{j,\sigma}n_{j,-\sigma} \right),
\end{align}
which has the gauge-transformed form of the Hamiltonian introduced in Eq. (1) of the main text. Note that $n_{j,\sigma}$, as well as all physical observables,  are gauge invariant, that is, they have the same expectation value in any reference frame. In fact, even the rotated-frame quasi-momentum is experimentally observable in time of flight by first adiabatically ramping up the Raman coupling and suddenly switching it off, which has the effect of projecting the dressed states on the dispersion of undressed spin states \cite{Lin-2011_sup}. Henceforth, here and in the main text as well, we will study the model in the gauge-rotated frame where translational invariance is manifest. 

\section{Dressed basis effective triangular ladder}\label{Ssec-triang_ladder_model}

In the absence of inter-atomic interactions, Hamiltonian \eqref{Seq_square_ham} is block diagonal in orthogonal quasimomentum  subspaces, $H_{\mathrm{n.i.}} = \sum_q H_q$, where  and $q = kd$ is the adimensional quasimomentum and $H_q$ reads
\begin{align}\label{Seq_square_hamq_long}
H_q = \sum_{\sigma=\pm 1/2} &\Bigg[ \left(4t\sin\left(\frac{\gamma}{2}\right)\sin(q)  + \delta\right)\sigma \tilde{a}_{q,\sigma}\dgr \tilde{a}_{q,\sigma} + \frac{\Omega}{2}\tilde{a}_{q,-\sigma}\dgr \tilde{a}_{q,\sigma} -\left( 2t\cos\left(\frac{\gamma}{2}\right)\cos(q) \right)\tilde{a}_{q,\sigma}\dgr \tilde{a}_{q,\sigma}\Bigg].
\end{align}
Here $\tilde{a}_{q,\sigma} = \frac{1}{\sqrt{L}} \sum_j a_{j,\sigma} \enum{-i q j}$ are the Fourier-transformed lattice modes. Equivalently, Hamiltonian \eqref{Seq_square_hamq_long} can be more compactly written in matrix form as 
\begin{equation}\label{SSeq_square_hamq}
H_q = \left(2t\sin\left(\frac{\gamma}{2}\right)\sin(q)+\frac{\delta}{2}\right)\sigma_z  -2t\cos\left(\frac{\gamma}{2}\right)\cos(q) + \frac{\Omega}2\sigma_x.
\end{equation}
Here, $\sigma_j$ are the spin-1/2 Pauli matrices, which act on the pseudospin space spanned by the coupled pair of internal states. In the diagonal or dressed basis, the noninteracting Hamiltonian can be rewritten as
\begin{equation}\label{Seq_square_ham_ni}
H_\mathrm{kin} = \sum_{q,m=\pm} \epsilon_{q,m}\tilde{b}_{q,m}\dgr\tilde{b}_{q,m},
\end{equation}
where the two dispersion bands are given by
\begin{equation}\label{Seq_bands}
\epsilon_{q,\pm} = \pm \frac{\Omega}2\sqrt{1 + (\tilde{q} + \tilde{\delta})^2} -2t\cos\left(\frac{\gamma}{2}\right)\cos(q),
\end{equation}
and where
\begin{equation}\label{Seq_bandmodes0}
\tilde{b}_{q,m'}\dgr = \sum_{m} \mathcal{U}_{m',m}(q) \tilde{a}_{q,m}\dgr,
\end{equation}
are the dressed modes of the noninteracting system. Here, we define $\tilde{q} \equiv  \frac{4t\sin(\gamma/2)\sin(q)}{\Omega}$ and $\tilde{\delta} \equiv \delta/\Omega$, and $\mathcal{U} = \enum{i\sigma_y\theta_q/2}$ is the unitary transformation that relates the bare and the dressed basis, with $\cos(\theta_q) = \frac{\tilde{q} + \tilde{\delta}}{\sqrt{1 + (\tilde{q} + \tilde{\delta})^2}}$, $0\le\theta_q\le\pi$. 

We now consider the regime where the rung tunneling $\Omega$ (the Raman coupling strength) is the dominant energy scale, and the two dispersion bands are well separated. We can then safely neglect the highest-band dressed states and truncate Hamiltonian \eqref{Seq_square_ham_ni} to
\begin{equation}\label{Seq_square_hamqtruncated}
H_\mathrm{kin} = \sum_{q} \epsilon_{q,-}\tilde{b}_{q,-}\dgr\tilde{b}_{q,-}.
\end{equation}
Finally, we introduce the inverse-Fourier-transformed truncated basis, the dressed modes
\begin{equation}\label{Seq_ftbandmodes}
b_n\dgr \equiv \frac{1}{\sqrt{L}}\sum_q\enum{-iqn}\tilde{b}_{q,-}\dgr = \frac{1}{L}\sum_{m,q, n'}\enum{iq(n'-n)}\mathcal{U}_{-,m}(q)a_{n',m}\dgr,
\end{equation}
and substitute \eqref{Seq_ftbandmodes} into \eqref{Seq_square_hamqtruncated}, which yields
\begin{equation}\label{Seq_square_ham_ni_truncated}
H_\mathrm{kin} \simeq \sum_n \sum_l t_{l} b_{n+l}\dgr b_{n},
\end{equation}
with
\begin{equation}\label{Seq_tunnelingcoeff}
t_{l} = \frac{1}{L}\sum_q\enum{-iql}\epsilon_{q,-}.
\end{equation}
By rewriting $\mathcal{U}$ explicitly in terms of $\tilde{q}$ and $\tilde{\delta}$, and considering the regime where $\tilde{q}\ll 1$, we have
\begin{equation}\label{Seq_rotation2}
\mathcal{U}_{-,\pm} = \pm\frac{1}{\sqrt{2}}\sqrt{1 \pm \frac{\tilde{q} + \tilde{\delta}}{\sqrt{1 + (\tilde{q} + \tilde{\delta})^2}}} = \pm \frac{1}{\sqrt{2}}\sqrt{1 \pm \frac{\tilde{\delta}}{\sqrt{1 + \tilde{\delta}^2}}}  + \frac{\tilde{q}}{ 2\sqrt{2}(1 +\tilde{\delta}^2)^{3/2}\sqrt{1 \pm \frac{\tilde{\delta}}{\sqrt{1 + \tilde{\delta}^2}}}} + O(\tilde{q}^2).
\end{equation}
By introducing \eqref{Seq_rotation2} into \eqref{Seq_ftbandmodes} and taking the limit $L\rightarrow \infty$, we obtain the following expression for the dressed mode operators $b_j\dgr$ in the bare basis $a_{j,m}\dgr$
\small
\begin{align}\label{Seq_truncatedmodes_1storder}
b_{j}\dgr = &\frac{1}{\sqrt{2}} \sqrt{1 + \frac{\tilde{\delta}}{\sqrt{1 + \tilde{\delta}^2}}} a_{j,+}\dgr - \frac{1}{\sqrt{2}}\sqrt{1 - \frac{\tilde{\delta}}{\sqrt{1 + \tilde{\delta}^2}}} a_{j,-}\dgr  \cr +&i\frac{ t\sin(\gamma/2)}{\sqrt{2}\Omega (1+\tilde{\delta}^2)^{3/2}}\left(\frac{a_{j+1,+}\dgr - a_{j-1,+}\dgr}{\sqrt{1 - \frac{\tilde{\delta}}{\sqrt{1 + \tilde{\delta}^2}}}} + \frac{a_{j+1,-}\dgr-a_{j-1,-}\dgr}{\sqrt{1 + \frac{\tilde{\delta}}{\sqrt{1 + \tilde{\delta}^2}}}}\right) + O\left((t/\Omega)^2\right).
\end{align}
\normalsize

Similarly, we can expand the expression for the effective tunneling coefficients $t_l$. From 
\begin{equation}\label{Seq_bands_truncated}
\epsilon_{q,-} = -\frac{\Omega}2\left(\sqrt{1+\tilde{\delta}^2} + \frac{\tilde{\delta}\tilde{q}}{\sqrt{1+\tilde{\delta}^2}} + \frac{\tilde{q}^2}{2(1 + \tilde{\delta}^2)^{3/2}}  + O(\tilde{q}^3)\right) -2t\cos\left(\gamma/2\right)\cos(q),
\end{equation}
and taking the limit $L\rightarrow \infty$, it follows that
\begin{align}\label{Seq_tunnelingcoeff_truncated}
t_l &= -t\Bigg[\left(\frac{\Omega}{2t}\sqrt{1+(\delta/\Omega)^2} + \frac{2t\sin^2(\gamma/2)}{\Omega(1+(\delta/\Omega)^2)^{3/2}}\right)\delta_{l,0}  +\left(\cos(\gamma/2) \mp i\frac{\delta\sin(\gamma/2)}{\Omega\sqrt{1+(\delta/\Omega)^2}}\right)\delta_{l,\pm 1} \nonumber \\
&-\frac{t\sin^2(\gamma/2)}{\Omega(1+\delta^2/\Omega^2)^{3/2}}\delta_{l,\pm 2} + O\left((t/\Omega)^2\right)\Bigg].
\end{align}

In this way, to first order in $t/\Omega$, the dressed Hamiltonian \eqref{Seq_square_ham_ni_truncated} can be written as
\begin{equation}\label{Seq_square_ham_ni_truncated_2}
H_\text{kin} = \sum_{i}( t_1 b_{i+1}^\dagger b_{i} + t_2 b_{i+2}^\dagger b_{i} + \text{H.c.})= -\sum_{i}( |t_1| e^{i\phi_1} b_{i+1}^\dagger b_{i} + |t_2| e^{i\phi_2} b_{i+2}^\dagger b_{i} + \text{H.c.}),
\end{equation}
where we drop the constant term $t_0N$ and only nearest and next-nearest neighbor couplings are kept. Hamiltonian \eqref{Seq_square_ham_ni_truncated_2} is equivalent to a triangular ladder with gauge invariant staggered flux $\Phi = \phi_2- 2\phi_1$, with $\phi_1 =\arg(-t_1)$ and  $\phi_2 = \arg(-t_2)$. Remarkably, in the chosen gauge the NN tunneling coefficient $t_1$ has a nonzero imaginary contribution that is linearly proportional to the Raman detuning $\delta$, see \eqref{Seq_tunnelingcoeff_truncated}. To first order in $\delta/\Omega$, we then have
\begin{equation}\label{Seq_staggered_fluxes}
\phi_1 = -\frac{\delta \tan(\gamma/2)}{\Omega} \text{ , } \quad \phi_2 = \pi . 
\end{equation}
Hence, at linear order in $\delta/\Omega$ around $\delta=0$, we can tune the staggered flux $\Phi$ while leaving $\vert t_1 \vert $ and $ \vert t_2 \vert$ unchanged. 
By setting $\delta=0$, the ladder is fully frustrated, with $\abs{\Phi} = \pi$, and we recover the expressions in equation (3) of the main text. 

Finally, if the interaction energy per particle is much smaller than the band gap, given by 
\begin{equation}
\Delta\epsilon = \operatorname{min}(\epsilon_{q,+} - \epsilon_{q',-}) = \Omega\left(1 - 4\cos(\gamma/2)\frac{t}{\Omega} + O\left( (t+\delta)^2/\Omega^2\right) \right),
\end{equation}
interatomic interactions can be treated within the lower band truncation. We consider SU(2) symmetric interactions
\begin{equation}\label{Seq_int_ham}
H_\mathrm{int} = \frac{U}{2}\sum_n N_n(N_n - 1),
\end{equation} 
where $N_n = \sum_{m} a_{n,m}\dgr a_{n,m}$. After truncating the expressions of $a_{n,m}$ and $a_{n,m}^\dagger$ to the lower band states, one has
\begin{equation}\label{Seq_Nn}
N_n \simeq \sum_{l,l'} C_{l,l'} b_{n+l}\dgr b_{n+l'},
\end{equation}
with
\begin{equation}\label{Seq_Ccoeffs}
C_{l,l^\prime} = \frac{1}{L^2}\sum_{q,q^\prime} \enum{i(ql-q^\prime l^\prime)}\sum_{m}\mathcal{U}_{-,m}(q)\mathcal{U}_{-,m}^*(q^\prime).
\end{equation}
From eqs. \eqref{Seq_rotation2} and \eqref{Seq_int_ham} to \eqref{Seq_Ccoeffs}, it follows that
\begin{align}\label{Seq_intham_approx}
H_\mathrm{int} \simeq  \frac{U}{2}\sum_n  b_{n}\dgr b_{n}(b_{n}\dgr b_{n}-1) + O((t/\Omega)^2). 
\end{align}
By combining the kinetic and interaction parts of the Hamiltonian \eqref{Seq_intham_approx} and \eqref{Seq_square_ham_ni_truncated_2}, and truncating to first order in $t/\Omega$, we recover the expression of the complete effective Hamiltonian (2) of the main text, which describes a triangular ladder configuration with tunable on-site interactions, tunnelings, and staggered flux, and is schematically represented in Fig.~1(a) of the main text.

The effective bosonic ladder can be further mapped to an \emph{XX} frustrated spin chain by considering hard-core bosons, with $U/\vert t_{1,2}\vert \rightarrow\infty$ and thus rendering double occupancies energetically forbidden. One can then identify $(b_{i}\dgr, b_j) \rightarrow (S_{i}^+, S_{j}^-)$, which leads to\begin{align}\label{Seq_hard-coreham}
H_{\bigtriangleup} \overset{U/t_l \rightarrow \infty}{\longrightarrow} \sum_j \left( t_1 S_{j}^{-}S_{j+1}^{+} + t_2 S_{j}^{-}S_{j+2}^{+} + {\rm H.c.} \right).
\end{align}
If we set $\delta=0$, or equivalently the effective staggered flux $\Phi = \pi$, such a HCB Hamiltonian describes a $t_1$-$t_2$ spin-$\frac{1}{2}$ \emph{XX}-chain 
\begin{equation}\label{Seq_hard_core_ham_delta0}
H_{\bigtriangleup}(\Phi = \pi) \overset{U/t_l \rightarrow \infty}{\longrightarrow} 2 t_1\sum_j \left( S_{j}^{x}S_{j+1}^{x} + S_{j}^{y}S_{j+1}^{y}\right) + 2 t_2 \sum_j \left( S_{j}^{x}S_{j+2}^{x} + S_{j}^{y}S_{j+2}^{y}\right),
\end{equation}
from which we infer the presence of frustrated regimes in the original semisynthetic square flux ladder.

\section{Current and density structures of the symmetry-broken phases}

\subsection{BOW phase as a vortex lattice insulator}

To gain insights on the nature of the BOW insulating phase of the effective triangular model back in the original square ladder, we now look at the properties of the ground state in the parameter regime that corresponds to the so-called Majumdar-Ghosh point of the \emph{XX} spin model \eqref{Seq_hard_core_ham_delta0}, that is,  at $t_2 = -t_1/2 > 0$ and $t_1/U \rightarrow 0$. There, the system is exactly solvable \cite{Majumdar_1970_sup}, with the ground state of the Hamiltonian being two-fold degenerate in the thermodynamic limit. Each solution spontaneously breaks the lattice translation symmetry of \eqref{Seq_hard_core_ham_delta0} and has periodicity two. Both are given by dimerized product states of triplet spin states defined on pairs of consecutive sites
\begin{equation}\label{Seq_Seq_MGeigenstates}
\ket{\psi_{\text{e,o}}} = \bigotimes_{j\in \text{even} / \text{odd}}\frac{\ket{\uparrow_{j}\downarrow_{j+1}}+ \ket{\downarrow_{j}\uparrow_{j+1}}}{\sqrt{2}} = \prod_{j\in \text{even} / \text{odd}}\frac{b_{j}\dgr + b_{j+1}\dgr}{\sqrt{2}}\ket{0}.
\end{equation}
Without loss of generality, in the following derivation we use $\ket{\psi_{\text{e}}}$. The ground state $\ket{\psi_{\text{e}}}$ can be written as 
\begin{equation}\label{Seq_MGeigenstatesDj}
\ket{\psi_{\text{e}}} = \prod_{j=0}^{L-1}D_j\dgr\ket{0},
\end{equation}
where we have defined the operators
\begin{equation}\label{Seq_Djoperators}
D_j\dgr = \frac{b_{2j}\dgr + b_{2j+1}\dgr}{\sqrt{2}},
\end{equation}
which fulfill $\left[D_j,D_k\dgr\right] = \delta_{j,k}$. At $\delta=0$, we conveniently reexpress the dressed mode operators \eqref{Seq_ftbandmodes}, truncated  to first order in $t/\Omega$, as
\begin{equation}\label{Seq_truncatedmodes_delta0}
b_{n}\dgr \simeq  \cos\alpha c_{n,-}\dgr +  i\sin\alpha \frac{ c_{n+1,+}\dgr - c_{n-1,+}\dgr}{\sqrt{2}},
\end{equation}
where we have defined $c_{n,\pm}\dgr = \frac{a_{n,+}\dgr \pm a_{n,-}\dgr}{\sqrt{2}}$ and $\alpha = \arctan\left(\frac{\sqrt{2}t \sin(\gamma/2)}{\Omega}\right)$.  By substituting the approximate expression \eqref{Seq_truncatedmodes_delta0} for the modes $b_j$ into \eqref{Seq_Djoperators}, we can write $D_j$ in the bare basis $\left\{c_{j,m}\dgr\ket{0}\right\}$ as
\begin{align}\label{Seq_Djapprox}
D_j\dgr  \simeq &\cos\alpha\frac{c_{2j,-}\dgr + c_{2j+1,-}\dgr}{\sqrt{2}} + i\sin\alpha \frac{c_{2j+2,+}\dgr + c_{2j+1,+}\dgr - c_{2j,+}\dgr - c_{2j-1,+}\dgr }{2}.
\end{align}

We now compute the vortex currents in the paired sites. We start by computing the expected value of the total rung current in the $2k$ plaquette, $j^{(r)}_{2k}$, defined as
\begin{align}
j^{(r)}_{2k} &= i\Omega\left( a_{2k,+} a_{2k,-}\dgr - a_{2k,+}\dgr a_{2k,-}\right) - i\Omega\left( a_{2k+1,+} a_{2k+1,-}\dgr - a_{2k+1,+}\dgr a_{2k+1,-}\right) \nonumber \\
&= -i\Omega\left( c_{2k,+} c_{2k,-}\dgr - c_{2k,+}\dgr c_{2k,-}\right) \nonumber +i\Omega\left( c_{2k+1,+} c_{2k+1,-}\dgr - c_{2k+1,+}\dgr c_{2k+1,-}\right).
\end{align}
By using expression \eqref{Seq_MGeigenstatesDj} and the commutation relations of operators $D_j$, it is easy to show that
\begin{align}\label{Seq_rungcurrent1}
 \bra{\psi_{\text{e}}} j^{(r)}_{2k}\ket{\psi_{\text{e}}} &= \bra{0} D_{k-1}D_{k}D_{k+1} j^{(r)}_{2k} D_{k-1}\dgr D_{k}\dgr D_{k+1}\dgr\ket{0} \nonumber \\
&=\bra{0} D_{k-1}D_{k}D_{k+1} \left[j^{(r)}_{2k}, D_{k-1}\dgr\right] D_{k}\dgr D_{k+1}\dgr \ket{0} \nonumber \\
&+\bra{0} D_{k-1}D_{k}D_{k+1} D_{k-1}\dgr \left[j^{(r)}_{2k}, D_{k}\dgr\right] D_{k+1}\dgr \ket{0} \nonumber \\
&+\bra{0} D_{k-1}D_{k}D_{k+1} D_{k-1}\dgr D_{k}\dgr \left[j^{(r)}_{2k}, D_{k+1}\dgr\right] \ket{0}.
\end{align}
Finally, inserting \eqref{Seq_Djapprox} into Eq. \eqref{Seq_rungcurrent1} and \eqref{Seq_legcurrent1} yields
\begin{align}
\bra{\psi_{\text{e}}} j^{(r)}_{2k}\ket{\psi_{\text{e}}} &= -\frac{\Omega}{\sqrt{2}}\sin\alpha\cos\alpha = -t \sin(\gamma/2) + O((t/\Omega)^3).
\end{align}

Similarly, we now compute the leg currents
\begin{equation}
 j^{\pm}_{k} = it\left(\enum{\mp i\gamma/2} a_{k,\pm}a_{k+1,\pm}\dgr - \enum{\pm i\gamma/2}a_{k,\pm}\dgr a_{k+1,\pm}\right).
\end{equation}
In the $c_{j,m}\dgr$ basis, the total leg current in the $2k$ plaquette reads
\begin{align}
j^{(l)}_{k} \equiv j^{-}_{k}-j^{+}_{k} = -t &  \left( c_{k,+} c_{k+1,+}\dgr +  c_{k,-} c_{k+1,-}\dgr   \right)\sin(\gamma/2) \cr
-it &\left( c_{k,+} c_{k+1,-}\dgr +  c_{k,-} c_{k+1,+}\dgr   \right)\cos(\gamma/2) + \mathrm{H.c.}
\end{align}
Again, we can write
\begin{align}\label{Seq_legcurrent1}
 \bra{\psi_{\text{e}}} j^{(l)}_{2k}\ket{\psi_{\text{e}}} &= \bra{0} D_{k-1}D_{k}D_{k+1} j^{(l)}_{2k} D_{k-1}\dgr D_{k}\dgr D_{k+1}\dgr\ket{0} \nonumber \\
&=\bra{0} D_{k-1}D_{k}D_{k+1} \left[j^{(l)}_{2k}, D_{k-1}\dgr\right] D_{k}\dgr D_{k+1}\dgr \ket{0} \nonumber \\
&+\bra{0} D_{k-1}D_{k}D_{k+1} D_{k-1}\dgr \left[j^{(l)}_{2k}, D_{k}\dgr\right] D_{k+1}\dgr \ket{0} \nonumber \\
&+\bra{0} D_{k-1}D_{k}D_{k+1} D_{k-1}\dgr D_{k}\dgr \left[j^{(l)}_{2k}, D_{k+1}\dgr\right] \ket{0}.
\end{align}
which after inserting \eqref{Seq_Djapprox} results in
\begin{align}
\bra{\psi_{\text{e}}} j^{(l)}_{2k}\ket{\psi_{\text{e}}} &= -t\cos(\gamma/2)\sqrt{2}\sin\alpha\cos\alpha -t\sin(\gamma/2)\left(\cos^2\alpha - \frac{1}{2}\sin^2\alpha \right) \cr &=-t \sin(\gamma/2)\left(1 + \frac{2t\cos(\gamma/2)}{\Omega}+ O((t/\Omega)^2) \right).\label{Seq_leg_current_VC}
\end{align}

To zeroth order in $(t/\Omega)$, we have $j_{2k}^{(l)} = j_{2k}^{(r)}$. This result can be extended to the $\ket{\psi_{\text{o}}}$ solution by just shifting each lattice site from $j$ to $j+1$. Therefore, the ground state solutions at the solvable Majumdar-Ghosh point correspond, in the original square ladder at quarter filling, to two vortex lattice states with vortex filling $\rho_{v} = 1/2$. The vortex current 
$j^{(c)}_{j} = j^{(l)}_{j}+j^{(r)}_{j}$ on the dimerized plaquettes is given by \begin{align}\label{eq_vortex_current_MG}
\bra{\psi_{\text{e}}} j^{(c)}_{2k}\ket{\psi_{\text{e}}} &= \bra{\psi_{\text{o}}} j^{(c)}_{2k+1}\ket{\psi_{\text{o}}} =  -2t \sin(\gamma/2) + O(t/\Omega). 
\end{align}
The first order correction to these solutions includes a small current $\propto t\cos(\gamma/2)/\Omega$ between the dimerized plaquettes that vanishes both for $t/\Omega \rightarrow 0$ and $\gamma \rightarrow \pi$. Notice that both solutions have a nonzero chiral current
\begin{equation}
j_c=\imath t\frac{2}{L}\sum_i\sum_{\sigma=\pm1/2}\sigma (e^{-\imath\gamma\sigma}a_{i+1,\sigma}^\dagger a_{i,\sigma}-e^{\imath\gamma\sigma}a_{i+1,\sigma} a_{i,\sigma}^\dagger).
\end{equation}
In particular, we have 
\begin{equation}\label{eq_jc_MG_point}
\bra{\psi_{\text{e,o}}} j_c \ket{\psi_{\text{e,o}}} = t\sin(\gamma/2)/2 + O(t/\Omega). 
\end{equation}
This net edge current is induced by the static gauge field that breaks the time-reversal symmetry in Hamiltonian (1) from the main text. These results have been derived assuming a HCB limit within the effective model, thus, to zeroth order in $t/U$, and to first order in $t/\Omega$. As we verified numerically, corrections to this leading behavior are tiny in the range of parameters considered in this work, where $U \gg \vert t_1\vert$ and $\Omega \gg t$, and a good qualitative agreement persists even for significantly smaller $\Omega$. The current distribution in the BOW state is schematically represented in  Fig.~1(b) from the main text, both in the effective model and in the semisynthetic flux ladder model. For the latter, we set $\Omega = 10t$, $\delta=0$ and $\gamma = 0.877\pi$, for which $t_2/\vert t_1\vert \sim 0.5$. The arrows representing the currents are scaled according to the values found numerically via DMRG simulations. The insulating nature of such a vortex phase can be inferred from the properties of the BOW phase of the \emph{XX} model, which is confirmed by numerical simulations of the full ladder model. Indeed, DMRG simulations reveal a finite single-particle excitation gap (see main text and Fig.~2(a)), and also the exponential decay of the one-body correlator $g^1$ with the total length of the ladder and a vanishing central charge when the system is extrapolated to the thermodynamic limit. In this way, we can identify the BOW phase of the lowest-band model with a vortex lattice insulating phase of vortex filling $\rho_{v} = 1/2$ (VL$_{1/2}$-MI) back in the original square flux ladder.

\begin{figure}[b]
\includegraphics[width=0.95\linewidth]{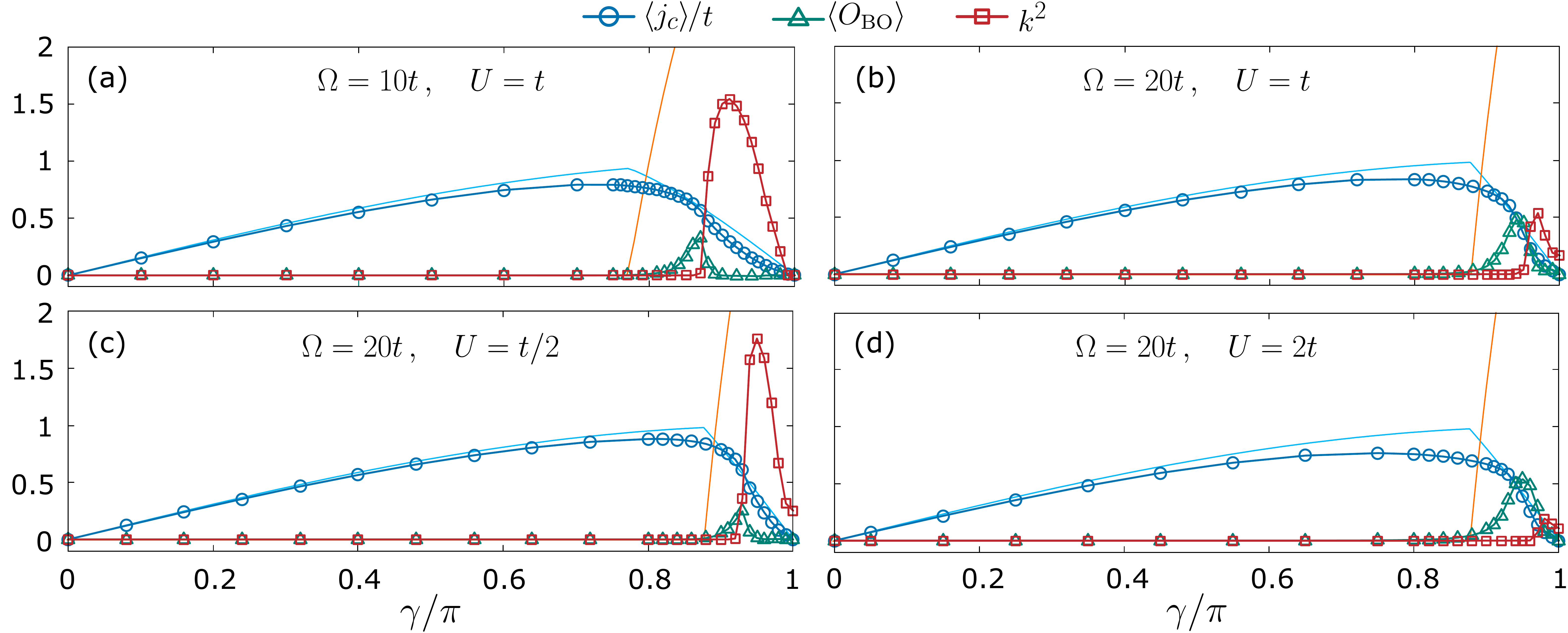}
\caption{Expected value of the chiral current ${\langle j_c\rangle}$ (blue circles), the bond-order operator $O_\mathrm{BO}$ (teal triangles) and the triangular ladder chiral correlation function $k^2(L)$ (dark red squares) as a function of $\gamma$ for the ground state of Hamiltonian (1) of the main text at $\rho = 1/4$. We set $\Omega = 10t$ and $U=t$ in (a), $\Omega = 20t$ and $U=t$ in (b), $\Omega = 20t$ and $U= 0.5t$ in (c), and $\Omega = 20t$ and $U=2t$ in (d). Cyan and orange thin solid lines show the expected values of the chiral current  and chiral correlation functions, respectively, in the noninteracting limit. Without interactions, no BOW exists and the bond-order operator $O_\mathrm{BO}$ is identically zero. In all cases, the values are obtained from a finite size extrapolation considering sizes up to $L=80$.}
\label{sfig_BO_k2_gamma}
\end{figure}

\subsection{The CSF as a biased ladder superfluid}

Similarly, the CSF phase of the effective triangular model has two degenerate solutions that spontaneously break the $Z_2$ parity symmetry of the Hamiltonian (2) of the main text, which corresponds in the semisynthetic square ladder to the spin-inversion symmetry, i.e. parity in the synthetic dimension, in addition to the U(1) phase symmetry. In each solution, local interatomic interactions favor the occupation of quasimomentum states around either the left or the right minimum of the band. 
In the corresponding regimes of semisynthetic flux ladder, the same behavior is expected. The current and density structure of the analogous phase back in the square ladder model is straightforwardly understood by taking the noninteracting limit of the two solutions $\ket{\psi_{\text{L,R}}}$, which are then simply given by the collective occupation of the band states located the left and right band minima, respectively. 

The single-particle band modes all have a vanishing rung current. For a bandstate at $q$, it follows
\begin{align}
\bra{0}\tilde{b}_{q,-}\, j_{k}^{(r)}\, \tilde{b}_{q,-}\dgr\ket{0} &=  \frac{i \Omega}{2L} \sum_{n n^{\prime} m m^{\prime}}\bra{0} \enum{-iq n^{\prime}} \enum{iq n} \mathcal{U}_{-,m^{\prime}}^*(q)\mathcal{U}_{-,m}(q) a_{n^{\prime},m^{\prime}}a_{k,-}a_{k,+}\dgr a_{n,m}\dgr\ket{0} + \mathrm{H.c.} \cr
&= -\frac{\Omega}L\operatorname{Im}\left(\mathcal{U}_{-,-}^*(q)\mathcal{U}_{-,+}(q) \right) = 0.
\end{align}
The leg current is given by
\small
\begin{align}\label{eq_leg_current_CSF}
\bra{0}\tilde{b}_{q,-} \, j_{k}^{\pm}\, \tilde{b}_{q,-}\dgr\ket{0} &=  \frac{i t}L  \sum_{n n^{\prime} m m^{\prime}}\bra{0} \enum{\mp i\gamma/2}\enum{-iq n^{\prime}} \enum{iq n} \mathcal{U}_{-,m^{\prime}}^*(q)\mathcal{U}_{-,m}(q) a_{n^{\prime},m^{\prime}}a_{k,\pm}a_{k+1,\pm}\dgr a_{n,m}\dgr\ket{0} \cr  & \quad + \mathrm{H.c.} \cr
&= -\frac{2}L\operatorname{Im}\left( t \enum{-i (q\pm \gamma/2)} \mathcal{U}_{-,\pm}^*(q)\mathcal{U}_{-,\pm}(q) \right) = \frac{2t}L\sin(q\pm \gamma/2)\abs{\mathcal{U}_{-,\pm}(q)}^2,
\end{align}
\normalsize
which yields a nonzero edge current
\begin{equation}
\bra{0}\tilde{b}_{q,-} \, j_{c}\, \tilde{b}_{q,-}\dgr\ket{0} = \frac{2t}L\sin(q+\gamma/2)\abs{\mathcal{U}_{-,+}(q)}^2 - \frac{2t}L\sin(q-\gamma/2)\abs{\mathcal{U}_{-,-}(q)}^2.
\end{equation}
At the same time, the band modes have a spin composition that depends on quasimomentum
\begin{equation}
\mean{m_z} = \bra{0}\tilde{b}_{q,-}\, \frac{n_{q,+}-n_{q,-}}2 \, \tilde{b}_{q,-}\dgr\ket{0} = \vert \mathcal{U}_{-,+}(q) \vert^2- \vert \mathcal{U}_{-,-}(q) \vert^2.
\end{equation}

If we now set $\delta=0$, where $\mathcal{U}_{-,\pm}(q) = \frac{1}{\sqrt{2}}\left(1\pm t\sin(\gamma/2)\sin(q)/\Omega\right) + O(\tilde{q}^2)$, see Eq. \eqref{Seq_rotation2}, we have
\begin{equation}
\mean{ j_{c}} = \frac{2t}L\left( \sin(\gamma/2)\cos(q) + \frac{2t}{\Omega}\sin(\gamma)\sin(q)^2 + O((t/\Omega)^2) \right),
\end{equation}
and
\begin{equation}
\mean{m_z} = \frac{2t\sin(\gamma/2)}{\Omega}\sin(q) + O((t\Omega)^2).
\end{equation}
For $\abs{\gamma} < \abs{\gamma_c} = \left \vert 4\tan^{-1}\left(\sqrt{\frac{\sqrt{(\Omega/t)^2+8^2}-8}{\Omega/t}}\right) \right \vert$, the lower band exhibits a single minimum at $q_m=0$, and the solutions correspond to the noninteracting limit of the Meissner phase: indeed, in this regime the chiral current increases with $\gamma$ as $\mean{ j_{c}} \propto \sin(\gamma/2)$, effectively screening the applied flux. Likewise, the solutions have a vanishing magnetization, that is, they evenly populate the two legs of the ladder.

For $\abs{\gamma}>\gamma_c$, the band splits into two degenerate minima at $q_{R,L}=\pm \cos^{-1}\left(\frac{\Omega\sqrt{1+(4t\sin(\gamma/2)/\Omega)^2}}{4t\tan(\gamma/2)}\right)$, and the chiral current for the corresponding states $\ket{\psi_{\text{R,L}}}$ starts to decrease with an increasing flux $\gamma$. In the absence of interactions, the change is nonanalytical in the thermodynamic limit, signaling a second order phase transition. This decrease in the flux screening is shared by the vortex superfluid phase of the flux ladder. However, in the CSF phase predicted by the effective model, the spontaneous occupation of either one of the two band minima $q_{R,L}$ prevents the formation of vortices. Instead, it gives rise to a population imbalance between the two legs, with $\mean{m_z}^{R,L} = \frac{2t\sin(\gamma/2)}{\Omega}\sin(q_{L,R})$. The CSF phase, thus, corresponds to a biased-ladder superfluid (BLP-SF) phase \cite{Wei_pra_BLP_theory_sup, Tokuno-njp-2014_sup, Greschner-prl-2015_sup, Greschner-2016_sup} back in the square flux ladder (see below). The currents and densities distributions in the CSF regime are schematically represented in Fig.~1(c) from the main text. 

The presence of interatomic interactions smooths out the single-minima to two-minima transition by favoring a spread of the momentum distribution that carries over into the two minima regime of the single-particle energy band \cite{DiDio_15_sup}. At particle filling $\rho = 1/4$ and in the presence of interatomic interactions, the Meissner to BLP transition is eventually lost in favor of the additional vortex lattice insulator phase in-between, that corresponds to the BOW dimer phase in the effective triangular model. Figure \ref{sfig_BO_k2_gamma} shows the expected values of the chiral current $\mean{j_c}$, $O_\mathrm{BO}$ and $k^2(L)$ for the ground state at $\rho = 1/4$ as a function of $\gamma$ for different values of $U$ and $\Omega$. The observables of the effective model $O_\mathrm{BO}$ and $k^2$ are computed back in the square ladder model by using the expressions (4) and (6) from the main text, respectively, which are derived below in this Supplemental Material. Convergence in the DMRG calculations is assessed by monitoring the eigenstate energy, with the state kept only once the energy updates are smaller than $10^{-8}t$. The same criterion is used in the simulations included in the main text. The light-colored blue and orange solid thin lines show the corresponding values obtained for the noninteracting flux ladder. At large $\gamma$, nonzero values of $O_\mathrm{BO}(L)$ and $k^2(L)$ for $L\rightarrow \infty$ signal the onset of the BOW and CSF phases, respectively. The BOW phase is favored at larger $\Omega$ and $U$. Contrarily, the CSF phase is suppressed by increasing atom-atom interactions and rung tunneling strengths. Still, the phase is expected to persist at fluxes $\gamma \sim \pi$ even for $\Omega \gg t$ and $U \gg \vert t_1 \vert, t_2$, as is predicted by the HCB limit of the lower-band model.

\section{Measuring the observables of the effective model}

\subsection{The effective chirality as inter-leg population imbalance}

By using equation \eqref{Seq_staggered_fluxes}, we can rewrite the effective chirality $k_j$ in terms of the susceptibility of the Hamiltonian against the interleg energy imbalance $\delta$. As shown in the previous sections, the parameters of the effective triangular model depend on $\delta$ only through the NN tunneling phase term $\phi_1$, with $\vert t_1 \vert$ and $t_2$ being constant to the order of approximation (see \eqref{Seq_tunnelingcoeff_truncated} and \eqref{Seq_staggered_fluxes}). With this in mind, it is easy to see that the effect of adding small local perturbations to $\delta$ simply results in a small spatial modulation of $\phi_1$, and we can write
\begin{equation}\label{Seq_staggered_flux_local}
\phi_1^{(j)} = -\frac{\delta_j \tan(\gamma/2)}{\Omega}, 
\end{equation}
where $\delta_j \sim \delta$ is the interleg energy imbalance at site $j$. In turn, since we have $\operatorname{Im}(t_1) = 0 + O(\delta/\Omega)$ (see \eqref{Seq_tunnelingcoeff_truncated}), in the vicinity of $\delta=0$ we can approximate $\operatorname{Im}( t_1 b_{j+1}\dgr b_j) \simeq  t_1 \operatorname{Im}(b_{j+1}\dgr b_j)$, and write the chirality $k_j$ as
\begin{equation}\label{Seq_chirality}
k_j = 4\operatorname{Im}( b_{j+1}\dgr b_j) \simeq \frac{1}{t_1} 4\operatorname{Im}( t_1 b_{j+1}\dgr b_j)   =  -\frac{2}{t_1}\frac{\partial H_{\bigtriangleup}}{\partial \phi_1^{(j)}}\Bigr|_{\substack{|t_1|,\phi_1^{(k)},t_2}},
\end{equation}
with $k\neq j$.
From \eqref{Seq_chirality} and \eqref{Seq_staggered_flux_local}, and given that $H_{\square} \simeq H_{\bigtriangleup}$ in the regimes considered, it follows that
\begin{equation}
k_j \approx \frac{2}{t_1} \frac{\partial \delta_j}{\partial \phi_1}\Bigr|_{\substack{|t_1|,\phi_1^{(k)},t_2}}\frac{\partial H_{\square}}{\partial \delta_j}\Bigr|_{\substack{|t_1|,\phi_1^{(k)},t_2}}  = -\frac{2\Omega}{t \sin(\gamma/2)}\frac{\partial H_{\square}}{\partial \delta_j}\Bigr|_{\substack{|t_1|,\phi_1^{(k)},t_2}}
\end{equation}
Note that since $\phi_1^{(k\neq j)}$, $|t_1|$ and $t_2$ are independent of $\delta_j$ at linear order, we have
\begin{equation}
\frac{\partial H_{\square}}{\partial \delta_j}\Bigr|_{\substack{|t_1|,\phi_1^{(k)},t_2}}=\frac{\partial H_{\square}}{\partial \delta_j}\Bigr|_{\substack{\Omega,\gamma, \delta_k}} = \frac{\partial H_{\square}}{\partial \delta_j}\Bigr|_{\substack{\Omega,\gamma, \delta_k}} = \sum_{\sigma} \sigma a_{j,\sigma}^\dagger a_{j,\sigma},
\end{equation}
and thus the chirality can be calculated over the original model with
\begin{equation}\label{Seq_current_mag_1}
k_j \approx -\frac{2\Omega}{t \sin(\gamma/2)} m_z^{(j)},
\end{equation}
where we have defined the local magnetization as
\begin{equation}\label{Se q_mean_pol}
m_z^{(j)} = \sum_{\sigma}\sigma a_{i,\sigma}^\dagger a_{i,\sigma}.    
\end{equation}

\subsection{Measuring the dimerization of the ground state}

The BOW phase is characterized by the spontaneous dimerization of the ground state, with two degenerate solutions that spontaneously break the translation symmetry of the Hamiltonian. Similarly, we can characterize the phase by the behavior of the system against the explicit breaking of the symmetry. Let us introduce an additional parameter $\Delta$ to the Hamiltonian 
\small
\begin{align}\label{Seq_square_ham_Delta}
H_{\square}(\Delta)=&\sum_{j, \sigma = \pm 1/2}\left(-t(1+\Delta(-1)^j)e^{-\imath\gamma\sigma}a_{j+1,\sigma}^\dagger +\frac{\Omega}{4} a_{j,-\sigma}^\dagger +\sigma\delta a_{j,\sigma}^\dagger \right)a_{j,\sigma} + \text{H.c.}
\nonumber\\
+&\sum_{j, \sigma}\left( \frac{U_{\sigma,\sigma}}{2}n_{j\sigma}(n_{j\sigma}-1)+\frac{U_{\sigma,-\sigma}}{2}n_{j,\sigma}n_{j-\sigma} \right).
\end{align}
\normalsize
Here, $\Delta$ fixes a relative dimerization of the longitudinal (spatial) tunneling in the square ladder. To first order, the addition of an infinitessimal value of $\Delta$ results into a dimerization of the effective triangular ladder
\begin{equation}
H_{\bigtriangleup}(\Delta)\simeq\sum_{j}(t_1(1+\Delta(-1)^j)b_j^\dagger b_{j+1} + t_2 b_j^\dagger b_{j+2} +\text{H.c.})+\frac{U}{2}\sum_{i}\tilde{n}_i(\tilde{n}_i-1),
\label{Seq_triang_Hamiltonian_Delta}
\end{equation}
that allows us to write the $O_\mathrm{BO}$ observable in the original basis as
\begin{equation}
O_\mathrm{BO} = \frac{1}{Lt_1}\frac{\partial H_{\bigtriangleup}}{\partial \Delta} \simeq -\frac{1}{L  t \cos(\gamma/2)}\frac{\partial H_{\square}}{\partial \Delta} = \frac{1}{L  t \cos(\gamma/2)} \sum_{j,\sigma} (-1)^j 2\operatorname{Re}\left(te^{-\imath\gamma\sigma} a_{j+1,\sigma}^\dagger a_{j,\sigma}\right).
\end{equation}

Furthermore, in the BOW phase the dimers in the effective triangular ladder correspond to vortices in the square ladder. With this in mind, it can be experimentally more convenient to probe the dimerization of the system simply by measuring the staggered current patterns. For instance, by measuring the response of the current 
\begin{equation}\label{Seq_staggerd_leg_current}
j_{sl} = \frac{1}{L} \sum_{j,\sigma} (-1)^j 4 \sigma \operatorname{Im}\left( te^{-\imath\gamma\sigma} a_{j+1,\sigma}^\dagger a_{j,\sigma}\right)
\end{equation}
against the variation of $\Delta$ around $\Delta=0$, the BOW (VL$_{1/2}$-MI) phase can be distinguished from the SF (M-SF) and CSF (BLP-SF) phases, as discussed in the main text. Similarly, one could equivalently measure the staggered pattern of the rung currents instead.

\section{Experimental considerations}

\begin{figure}[t]
\includegraphics[width=0.5\linewidth]{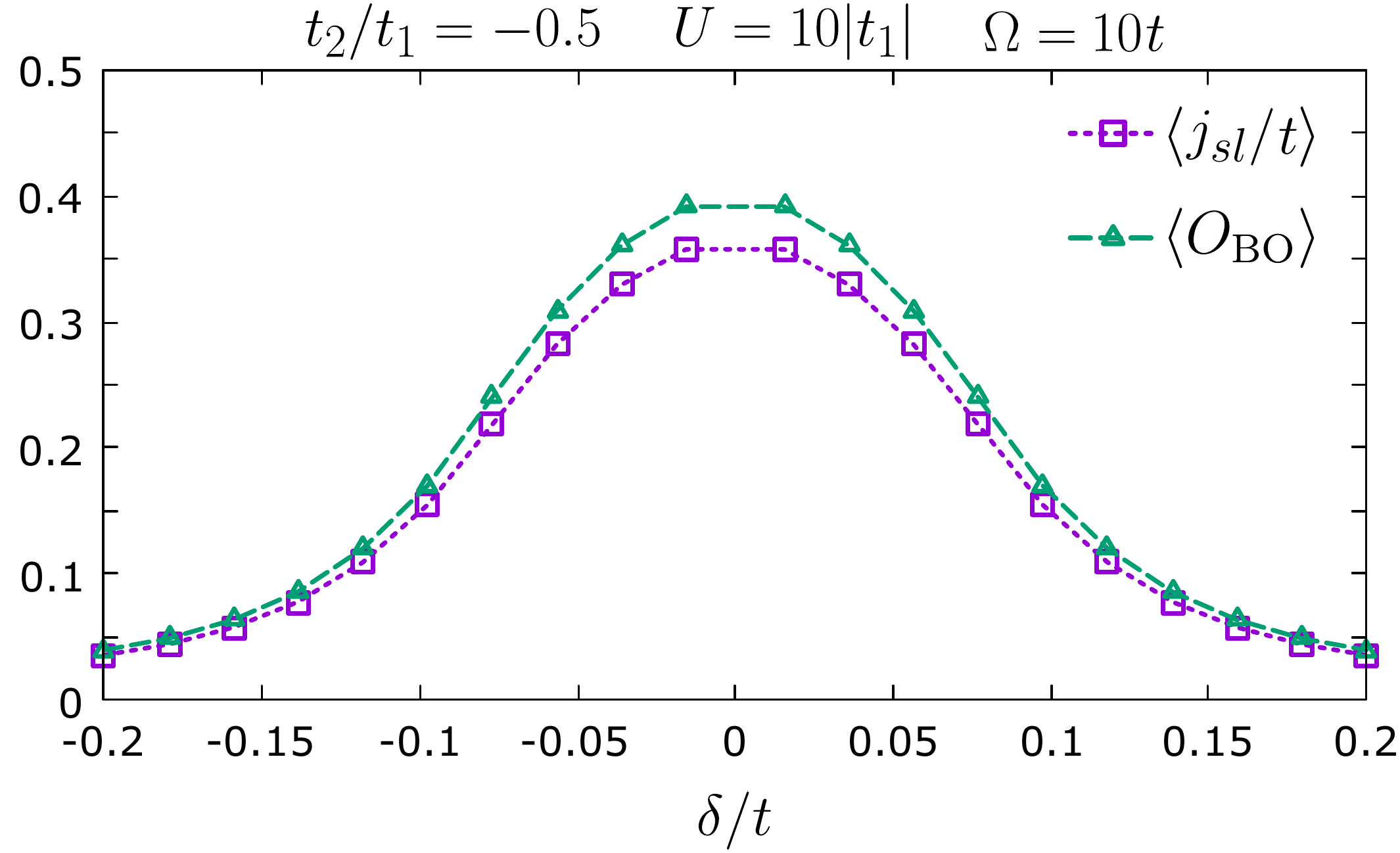}
\caption{Expected value of staggered leg current $j_{sl}$ (purple square dots, see Eq.\eqref{Seq_staggerd_leg_current}) and $O_\mathrm{BO}$ (green triangle dots) of the ground state of Hamiltonian (1) of the main text as a function of $\delta$, with $\Omega=10t$ and $\gamma$ and $U$ adjusted so that $U = 10 \vert t_1 \vert$ and $t_2/\vert t_1\vert = 0.5$. All quantities are extrapolated to the thermodynamic limit by considering system sizes up to $L= 80$.}
\label{sFig_OBO_v_delta}
\end{figure}

\emph{Atomic species}. The semisynthetic bosonic flux ladder discussed in the main text can be implemented with essentially any optically-coupled bosonic atomic species, including Raman-coupled $^{87}$Rb. In this work, we focus instead on $^{41}$K because the energy splitting of states $\ket{\uparrow}\equiv\ket{F=1, m_F = -1}$ and $\ket{\downarrow}\equiv\ket{F=1, m_F = 0}$ at a magnetic field $B_0=338.4$ G is to first order insensitive to magnetic field fluctuations, allowing for an excellent control of the two-photon Raman detuning $\delta$ \cite{Froelian-thesis_sup} to which the BOW phase is quite sensitive, see \fref{sFig_OBO_v_delta}. Moreover, at $B_0$ the scattering lengths characterizing the interatomic interactions have an almost perfect SU(2) character ($a_{\uparrow\uparrow}=60.9a_0$, $a_{\downarrow\downarrow}=61a_0$, and $a_{\uparrow\downarrow}=60.7a_0$), as assumed in the main text .

\emph{1D optical lattice}. To realize the real-space dimension of the semisynthetic ladder, the atoms must be subjected to a 1D optical lattice. We select a short lattice wavelength $\lambda_{L,s} = 532$ nm and a retro-reflected beam configuration in order to minimize the lattice spacing $d_s=\lambda_{L,s}/2$, since this enhances the lattice recoil energy $E_{L,s}=h^2/(8 m d_s^2)$ and thus the characteristic energy scales of the effective model (see below). Here $h$ is Planck's constant and $m$ the mass of $^{41}$K. For a lattice depth $V_{L,s} = 5E_{L,s}$, the tunneling rate in the square ladder configuration is $t/h \simeq 1.1$ kHz. Moreover, the atoms need to be tightly confined along the transverse directions to enter the one-dimensional regime. This can be achieved using two additional retro-reflected optical lattices along the perpendicular directions with wavelength $\lambda_{L,l}=1064$ nm and depth $V_{L,l} \sim 46E_{L,l}$, which will create an array of decoupled one-dimensional lattice systems. In these conditions, the interaction energy becomes $U/h\sim 2.1$ kHz and its value can be finely controlled by adjusting the transverse confinement. 

\emph{Raman coupling}. The tunneling along the synthetic dimension is provided by coupling states $\ket{\uparrow}$ and $\ket{\downarrow}$ via two-photon Raman transitions with Rabi coupling strength $\Omega$, which provide easy access to the regime $\Omega=10t$. We select the potassium tune-out value $\lambda_R = 769$ nm for the Raman beams. This ensures a favorable ratio of the Rabi coupling to the inelastic photon scattering rate by spontaneous emission from the Raman beams -- which leads to heating -- and does not create an additional scalar potential on the atoms. To obtain a synthetic flux $\gamma\sim\pi$, each of the Raman beams must form an angle $\theta_R\sim45 ^{\circ}$ with the lattice beams, since $\gamma/(2\pi)=(\lambda_{L,s}/\lambda_R)\cos{\theta_R}$ \cite{Celi-2014_sup}. The exact flux value can be adjusted by controlling $\theta_R$ (see below).

\emph{State preparation  with  an  optical  superlattice.} The phases investigated in this work appear at half filling in the effective triangular lattice, which corresponds to quarter filling in the original square flux ladder. Disregarding the synthetic spin dimension, the required filling in the original 1D lattice is therefore $1/2$ particle per site. To obtain this value, a robust strategy consists on preparing first a unity-filled Mott insulator in a $\lambda_{L,l}=1064$ nm retro-reflected lattice, where the lattice spacing is twice $d_{s}$, before transferring the atoms to the final $\lambda_{L,s}=532$ nm lattice. This will result in a half-filled 1D lattice and, after turning on the synthetic Raman tunneling, yield the required $\rho=1/4$ filling in the synthetic square flux ladder.

\emph{Experimental scales.} The BOW phase appears in the strongly interacting regime of the effective model. Realizing it requires both a large band gap and a small bandwidth, and thus large values of $\Omega$. Specifically, setting as above $t/h=1.1$ kHz and $\Omega=10t$, the tunneling in the effective model becomes $\vert t_1/h \vert \sim 213$ Hz and the corresponding interaction energy $U/h\sim 2.1$ kHz (which remains unchanged) corresponds to $U/|t_1|\sim 10$. To access the BOW regime, i.e. $\vert t_2/t_1 \vert \sim 1/2$, a flux $\gamma \sim 0.88\pi$ is then required, which can be obtained by setting the an angle of the Raman beams with respect to the lattice beams to $\theta_R\sim50.5^{\circ}$. For $\Omega = 10t$ the inelastic photon scattering from the Raman beams due to spontaneous emission we expect a gas lifetime of $\sim140$ ms \cite{Wei_2013_sup, Froelian_2022_sup}, which is about 30 times larger than the characteristic tunneling timescales of the effective model. 
Finally, we note that the main factor limiting the experimental system size is the harmonic confinement created by the transverse optical lattice beams, which could be compensated using a digital micromirror device \cite{Wei_Science_2022_sup}. By doing so, system sizes comparable to the ones used in our DMRG simulations could be reached in the experiment.

\emph{Detection.} The symmetry breaking measurements of Fig.~3 of the main text are straightforward to implement experimentally. On the one hand, the magnetization vs. detuning curve can be simply determined by measuring the leg populations -- which in a semisynthetic ladder correspond to the spin populations -- through Stern-Gerlach separation in a time-of-flight experiment. On the other hand, for the staggered leg current vs. lattice dimerization curve, the dimerizing potential can be obtained using the same $1064$ nm optical superlattice as for the preparation scheme, and the staggered leg current can be determined from the plaquette dynamics as in previous experiments \cite{Atala_2014_sup}. Thus, our proposal should be immediately accessible to current experiments.

\end{document}